# All AMMs are CFMMs. All DeFi markets have invariants.
# A DeFi market is arbitrage-free if and only if it has an increasing invariant

Roger Lee[*]

This version: December 29, 2023

**Abstract**

In a universal framework that expresses any market system in terms of state transition rules, we prove that every DeFi market system has an invariant function and is thus by definition a CFMM; indeed, all automated market makers (AMMs) are CFMMs.

Invariants connect directly to arbitrage and to completeness, according to two fundamental equivalences. First, a DeFi market system is, we prove, arbitrage-free if and only if it has a strictly increasing invariant, where *arbitrage-free* means that no state can be transformed into a dominated state by any sequence of transactions. Second, the invariant is, we prove, unique if and only if the market system is *complete*, meaning that it allows transitions between all pairs of states in the state space, in at least one direction.

Thus a necessary and sufficient condition for no-arbitrage (respectively, for completeness) is the existence of the increasing (respectively, the uniqueness of the) *invariant*, which, therefore, fulfills in nonlinear DeFi theory the foundational role parallel to the existence (respectively, uniqueness) of the pricing measure in the Fundamental Theorem of Asset Pricing for linear markets.

Moreover, a market system is recoverable by its invariant if and only if it is complete; and in all cases, complete or incomplete, every market system has, and is recoverable by, a multi-invariant. A market system is arbitrage-free if and only if its multi-invariant is increasing.

Our examples illustrate (non)existence of various specific types of arbitrage in the context of various specific types of market systems – with or without fees, with or without liquidity operations, and with or without coordination among multiple pools – but the fundamental theorems have full generality, applicable to any DeFi market system and to any notion of arbitrage expressible as a strict partial order.

---

[*]Department of Mathematics, University of Chicago. `rogerlee@math.uchicago.edu`

# 1  Introduction

Decentralized finance (DeFi) systems offer financial services via distributed ledger technologies, involving software known as smart contracts, on blockchains such as Ethereum. These services include crypto analogues of intermediation functions traditionally offered by centralized financial institutions such as banks and exchanges. For instance, an automated market maker (AMM) can serve as a decentralized exchange (DEX) by holding assets in liquidity pools, against which traders may transact, by choosing from among the trades offered by the AMM, which algorithmically determines what output bundle to offer in exchange for each possible input bundle [17, 8].

Depositors of assets into DeFi protocols face risks including the exploitation of the system by arbitrageurs conducting economic attacks to drain those assets, by using or abusing the operations offered by the system. In such scenarios, some of DeFi's hallmark features – such as its accessibility, composability, and disintermediation of humans (including risk managers and regulators) – become potential vulnerabilities, amplifying the feasibility and/or the impact of an attack. Thus the question of which markets do or do not admit *arbitrage* – already crucial in traditional finance – takes on additional significance in DeFi.

In a universal framework that expresses any market system in terms of state transition rules, this paper establishes the equivalence between, on one hand, the no-arbitrage (and completeness) properties of DeFi systems[1], and on the other hand, the existence (and uniqueness) of increasing *invariant* functions on the state space. The concepts of invariants, arbitrage, and completeness, therefore, constitute an interconnected *triad*. Let us outline, in turn, each member of the triad.

The menu of transactions offered by an AMM or a DeFi system may or may not be documented in terms of an *invariant* function that invariably stays unchanged after each reversible transaction, and invariably becomes strictly larger after each irreversible transaction. For instance, non-DEX DeFi services, and the early versions of AMMs for prediction markets [15], do not typically reference invariants in their specifications; on the other hand, a DEX typically specifies some of (but not all of) its operations in terms of invariants. In any case, irrespective of how a DeFi system describes its operations, the "zeroth" theorem is that every AMM and every DeFi system does indeed, we prove, *admit* an invariant function valid on *all* operations of the system. Therefore, every AMM is a *CFMM* (defined[2] as a market system that has an invariant). The zeroth theorem makes an order-continuity assumption in defining AMM, but that assumption is satisfied by all truly *automated* systems in practice.

Invariants relate directly to absence of arbitrage, according to the first fundamental theorem: a DeFi system is arbitrage-free if and *only* if it admits an *increasing invariant*. In other words, the increasing CFMMs are, we prove, exactly the *arbitrage-free* systems, in the sense that no sequence of operations can transform the system to a "worse" or "dominated" state. The definition

---

[1] *DeFi system* and, synonymously, *DeFi market* and *DeFi market system*, are defined in Section 3.3.

The *system* terminology emphasizes our flexibility to model a network/ecosystem of primitives.

[2] *CFMM* is defined in Definition 2.8, consistently with common language, as discussed in Remark 2.21, fifth point.

of "dominated", and therefore the notion of arbitrage, is fully flexible, provided that the invariant's monotonicity is understood in the same sense. The scope of "operations" is also fully flexible; it comprises every transition rule in the system, and that ruleset is adjustable according to whatever functionality the designer/auditor/researcher wishes to analyze – swaps, liquidity operations, donations, or anything else. The first fundamental theorem makes order-continuity and "giftability" assumptions in defining a DeFi system, but those conditions hold in all standard blockchain wallets; and the "if" direction does not assume even those conditions.

Invariants relate directly to *completeness*, according to the second fundamental theorem: a CFMM is complete if and only if its invariant is unique. More precisely, if for every pair of states in the state space, at least one member of the pair can be reached from the other member, then the CFMM has only one invariant function (modulo strictly increasing mappings between invariants); otherwise, the CFMM admits multiple invariants.

These equivalences constitute the nonlinear AMM / DeFi analogue of the set of linear-market equivalences known as [16] the *Fundamental Theorem of Asset Pricing* (FTAP), which establishes, loosely speaking, that a linear[3] market is arbitrage-free[4] if and only if it admits a pricing measure (versions of which are known, in various manifestations, as a [local] martingale measure, or state prices, or risk-neutral probabilities), and that an arbitrage-free linear market is complete[5] if and only if it admits a unique pricing measure. The classical FTAP does not apply to the nonlinear AMMs of DeFi, but by introducing DeFi-appropriate notions of arbitrage and completeness, analogues of the Fundamental Theorem emerge in the DeFi space (Theorems 7.1,7.2), as described in the triad discussion above – where the increasing *invariant* replaces the *pricing measure* as the foundational structure enforcing consistency relationships across all trades, whose existence is equivalent to absence of arbitrage, and whose uniqueness is equivalent to market completeness.

Complementary to *geometric* approaches [3, 4], this paper takes a *topological* and *poset-theoretic* approach; insights from geometric features such as convexity and gradients are not explored here; nor is existence of such features assumed here. Complementary to insights from other axiomatic analyses of AMMs, this paper's approach of examining the features of the *triad* (arbitrage-invariance-completeness) in *general* market systems leads us to axiomatizations different from approaches [14] which restrict to within the REMM subclass of market systems, and/or which target features outside of the triad (such as the features examined in [6], which overlap with only part of one of the five implications contained in our triad's fundamental Theorems).

Our approach results in contributions including: universality of the framework, minimality of assumptions, and applicability of conclusions. First, the *framework is universal*, in the sense that the poset-theoretic approach does not bind us to any one notion of arbitrage, but rather allows arbitrage to be defined by any strict partial order on the state space; and also universal in the sense that the framework includes any system expressible in terms of state-transition rules – which includes every machine-implementable protocol or network of protocols.

---

[3]in the sense that the total price of any bundle of traded assets is linear in quantity. We do not assume linear markets and do not use the classical FTAP.

[4]in a classical finance-theory sense that will not need to be recalled in our minimalistic framework, where the appropriate notion of arbitrage comes directly from the theory of partially ordered sets.

[5]in a classical finance-theory sense that will not need to be recalled in our minimalistic framework, where the appropriate notion of completeness comes directly from preference theory rather than finance theory.

Second, the *assumptions are minimal*, in that our focus on the connections within the arbitrage-invariance-completeness triad requires, for some results, *nothing* outside of the triad to be assumed, and for other results, only mild assumptions that hold in all ordinary blockchain wallets.

Third, the conclusions have *robust applicability*, in that our examples prove robustly the absence of arbitrage under a variety of realistic conditions: robust to the possibility that exploiters may use not only swaps but also liquidity operations to attack a system; robust to the presence of fees – not only the fees accruing *to* liquidity providers (LPs) but also admin fees extracted *from* LPs by a protocol via dilutive minting of liquidity tokens; and robust to threats against not just an individual pool, but also weighted combinations of all pools in a coordinated multi-pool *system*. To highlight one example, given any CFMM that does not support liquidity addition/removal but that has an increasing invariant (and, therefore, has guaranteed unexploitability of the operations that it supports), we build an increasing invariant on an augmented market incorporating liquidity operations, and thereby extend the guaranteed unexploitability to all combinations of the original operations and/or liquidity operations (properly designed). To spotlight another example, we answer rigorously, in what sense a Uniswap v3-style network of liquidity pools is arbitrage-free.

## 2 Market systems

**Definition 2.1.** A *market system* on a *state space* $X$ is a function $M : X \longrightarrow 2^X$, where $2^X$ denotes the set of subsets of $X$.

Our use cases will take $X \subseteq \mathbb{R}^n$, but our theory allows general sets $X$. Intuitively $M$ maps each state $x \in X$ to the set $M(x)$ of states attainable by a single transaction starting from $x$, where the single transaction is any member of the "menu" of valid transactions offered by the market system.

Let $\mathbb{R}_+ := [0, \infty)$ and $\mathbb{R}_{++} := (0, \infty)$.

**Example 2.2** (Constant-product market maker that takes a 0.3% fee)**.** Let $X = \mathbb{R}_{++}^2$ and

$$M(x_1, x_2) := \{(y_1, y_2) \in X : (y_1 - \phi_1)(y_2 - \phi_2) = x_1 x_2 \text{ where } \phi_j := 0.003 \max(y_j - x_j, 0)\}. \quad (2.1)$$

Here $x_1$ and $x_2$ are quantities (known as balances or reserves) of "asset 1" and "asset 2" in a pool. This market system supports swaps into any state $(y_1, y_2)$ such that the new balances, minus 0.3% of the increase in the incoming asset, have the same product as the balances in the previous state.

**Example 2.3** (Constant-product, with fees, liquidity operations, and gifts)**.** To bring liquidity operations and gifts/donations (which Uniswap v2 accepts [1]) into Example 2.2, let $X = \mathbb{R}^3_{++}$ and

$$M(x_1, x_2, x_\ell) := \{(y_1, y_2, y_\ell) \in X : (y_1 - \phi_1)(y_2 - \phi_2) = x_1 x_2 \text{ and } x_\ell = y_\ell\} \tag{2.2}$$
$$\cup \{\alpha(x_1, x_2, x_\ell) : \alpha > 0\} \cup \{(y_1, x_2, x_\ell) : y_1 > x_1\} \cup \{(x_1, y_2, x_\ell) : y_2 > x_2\},$$

in the style of Uniswap v2 (but without roundoff). Here $x_\ell$ and $y_\ell$ are the pre-transaction and post-transaction total supply of liquidity provider shares (*LP shares* or *LP tokens*) that represent ownership stakes in the two-asset pool. The second set in the union allows multiplying the state vector $(x_1, x_2, x_\ell)$ by a scalar $\alpha$, in order to provide (if $\alpha > 1$) or remove (if $\alpha < 1$) liquidity. The third and fourth sets allow donations of asset 1 or 2 into the pool. The fee $\phi$ is defined as in (2.1).

In general, the components of the state vectors in $X$ can include any variable tracked by the system. Examples may include the balances of any number of assets (in each pool within any set of pools, in each protocol within any ecosystem of DeFi protocols), quantities of liquidity tokens, volumes, fee totals, time stamps, oracle data such as prices and market depth metrics, functions of the *history* or *path* of any variable, such as lagged or time-averaged or differenced variables, and wallet-level variables for individual users.

Indeed Definition 2.1 is universal, in that it encompasses every conceivably implementable combination of DeFi primitives, because on a coding level, any protocol (or network of protocols) operates by transforming state variables.

**Definition 2.4.** Given a market system $M$ on $X$, define $M_\infty : X \to 2^X$ by

$$M_\infty(x) := \bigcup_{j=0}^{\infty} M_j(x) \tag{2.3}$$

where $M_0(x) := \{x\}$ and $M_{j+1}(x) := \{z : z \in M(y) \text{ for some } y \in M_j(x)\}$.

Thus $M_\infty(x)$, which we may call the *reachable* set from $x$, is the set of states attainable by any finite sequence of 0 or more transactions offered by $M$, starting from state $x$. Likewise, $M_1(x)$ (equivalently, $M(x)$) is the set of states attainable in 1 step (by a single transaction) from $x$.

**Example 2.5.** Let $X = \mathbb{R}^3_{++}$ and $M$ and $\phi$ be as in the stylized Uniswap v2 Example 2.3. Define

$$M'(x_1, x_2, x_\ell) := \{\alpha(x_1, x_2, x_\ell) : \alpha > 0\} \cup \{(y_1, y_2, y_\ell) \in X : (y_1 - \phi_1)(y_2 - \phi_2) \geq x_1 x_2 \text{ and } x_\ell = y_\ell\}.$$

It is easy to verify that $M'_\infty = M_\infty$. Consequently, for any analysis of properties that depend only on the reachable sets, it may be convenient to use a formulation such as $M'$ which allows combined operations of swapping and gifting (donating), rather than (2.2) which separates the operations.

**Example 2.6.** Let $X = \mathbb{R}^3_{++}$ and $M$ and $\phi$ be as in Example 2.3. Let

$$M''(x_1, x_2, x_\ell) := \left\{(y_1, y_2, y_\ell) \in X : (y_1 - \phi_1)(y_2 - \phi_2) \geq x_1 x_2 \text{ and } y_\ell = \frac{6\sqrt{y_1 y_2}}{5\sqrt{y_1 y_2} + \sqrt{x_1 x_2}} x_\ell\right\}$$
$$\cup \{\alpha(x_1, x_2, x_\ell) : \alpha > 0\}. \tag{2.4}$$

In contrast to Example 2.5 where the full 0.3% fee goes to the pool, this example redirects one-sixth of the 0.3% as a "protocol fee" or "admin fee" (which Uniswap v2 provides the ability to activate via governance, and which Sushiswap does activate), by minting liquidity shares, as specified in [1]. The destination of the minted admin fee is a treasury wallet of no concern to us in this example (which takes the standpoint of the liquidity providers), so (2.4) simply updates $y_\ell$ to recognize the dilution of liquidity shares. If however some other use case arises, in which the admin fee's destination plays some role in whatever notion of arbitrage is to be studied, then this framework can accommodate extra state variables to track any other relevant wallets.

**Definition 2.7.** Define the preorder (reflexive transitive relation) $\precsim_M$ induced by a market system $M$ on a state space $X$ by

$$x \precsim_M y \text{ if and only if } y \in M_\infty(x). \tag{2.5}$$

We may abbreviate $\precsim_M$ as $\precsim$ or as $\to$ when there is no ambiguity about $M$, as we do in (2.6,2.7).

Define the equivalence relation $\sim_M$ induced by a market system $M$ by

$$x \sim_M y \text{ if and only if } x \precsim y \text{ and } y \precsim x. \tag{2.6}$$

Define the strict preorder or strict partial order (asymmetric transitive relation, or equivalently: irreflexive transitive relation) $\prec_M$ induced by a market system $M$ by:

$$x \prec_M y \text{ if and only if } x \precsim y \text{ and } not \ y \precsim x. \tag{2.7}$$

We may abbreviate $\prec_M$ and $\sim_M$, respectively, as $\prec$ and $\sim$, when there is no ambiguity about $M$.

In other words, $x \precsim y$ or $x \to y$ means that $y$ is "reachable" from $x$ in the sense that the system allows state $x$ to be transformed into state $y$ in some finite number of transactions. If $x \precsim y$, two subcases exist: $x \prec y$ means the transformation is irreversible, in that $x$ is not reachable from $y$; while $x \sim y$ means the transformation is reversible, in that $x$ and $y$ can be reached from each other.

## 2.1 Invariants

**Definition 2.8.** Let $M$ be a market system on a state space $X$, and let $\prec$ and $\sim$ denote its induced strict preorder and equivalence relation respectively. If there exists $K : X \to \mathbb{R}$ satisfying

$$K(x) < K(y) \quad \text{for all } x, y \in X \text{ such that } x \prec y, \tag{2.8}$$

$$K(x) = K(y) \quad \text{for all } x, y \in X \text{ such that } x \sim y, \tag{2.9}$$

then we say that $K$ is an *invariant* of $M$, and that $M$ is a *CFMM*, and that $M$ has invariant $K$.

Equivalently, $M$ has invariant $K$, if $K(x) \leq K(y)$ for all $x \precsim y$, and $K(x) < K(y)$ for all $x \prec y$.

Equivalently, $M$ has invariant $K$, if $K(x) \leq K(y)$ for all $x \precsim y$, with equality *only if* $x \sim y$. To be precise, the "equality only if" stipulation is that if $x \precsim y$ and $K(x) = K(y)$ then $x \sim y$.

The fifth point of Remark 2.21 further discusses this definition.

For instance, the market system in Example 2.2 has invariant $K(x_1, x_2) = x_1 x_2$; the product $x_1 x_2$ invariably stays unchanged after each reversible transaction, and invariably becomes strictly larger after each irreversible transaction. Each of the market systems in Examples 2.5 and 2.6 has invariant $K(x_1, x_2, x_\ell) = x_1 x_2 / x_\ell^2$, to be discussed in Proposition 4.4 and Example 4.5.

**Definition 2.9.** A market system on $X$ is said to be *complete* if for all $x, y \in X$, we have

$$x \precsim y \text{ or } y \precsim x. \tag{2.10}$$

In other words, a market system is complete if it induces a *total* preorder.

**Definition 2.10.** We say that an invariant $K$ *recovers* a market system $M$, or that $M$ is *recoverable* by invariant $K$, if for all $x, y \in X$, we have

$$K(x) \leq K(y) \quad \text{if and only if } x \precsim y. \tag{2.11}$$

Equivalently,

$$K(x) < K(y) \quad \text{if and only if } x \prec y, \tag{2.12}$$
$$K(x) = K(y) \quad \text{if and only if } x \sim y. \tag{2.13}$$

In other words, recovery means that the converses of (2.8) and (2.9) also hold.

**Proposition 2.11.** *Let $M$ be a market system with an invariant $K$. Then $M$ is complete if and only if $K$ recovers $M$.*

*Proof.* For each $x, y \in X$, one of $K(x) < K(y)$ or $K(x) = K(y)$ or $K(y) < K(x)$ holds, so if $K$ recovers $M$, then for each $x, y \in X$, one of $x \prec y$ or $x \sim y$ or $y \prec x$ holds, hence $M$ is complete. Conversely, if $M$ is complete, then for any $x, y \in X$ if not $x \prec y$ then either $x \sim y$ or $y \prec x$ hence $K(y) \leq K(x)$, which verifies (2.12), and likewise if not $x \sim y$ then either $x \prec y$ or $y \prec x$ hence $K(y) \neq K(x)$, which verifies (2.13) and thus recovery by $K$. □

## 2.2 Arbitrage

To say that a market system admits arbitrage means, intuitively, that some sequence of transactions can transform some state into a "worse" or "dominated" state. Thus the definition of arbitrage depends on the definition of *worse/dominated*. This paper's general framework offers freedom for anyone to specify the sense in which a better state "dominates" a worse state, depending on whatever one wishes to regard as arbitrage in the context of the application at hand.

The particular examples of arbitrage chosen in this paper take the standpoint of the market system (or more precisely: passive depositors of liquidity into the market system), and ask whether the system can be exploited by some combination of transactions. Our examples here will not take the standpoint of the attacker; they will not directly ask whether an attacker can make arbitrage profits, but rather whether the system can suffer arbitrage losses.

**Definition 2.12.** A *dominance order* D is a strict partial order on a state space $X$.

**Definition 2.13.** Consider the following examples $\{\mathsf{D}_\Pi, \mathsf{D}_{\Pi/\mathsf{L}}\}$ of dominance orders.

Define the "Pareto" or "product" order $\mathsf{D}_\Pi$ on $X \subseteq \mathbb{R}^n$ by

$$x \; \mathsf{D}_\Pi \; y \text{ if and only if } x_i \geq y_i \text{ for every } i = 1, \ldots, n, \text{ and } x \neq y. \tag{2.14}$$

and define the "Pareto per-LP-share" order on $X \subseteq \mathbb{R}^n \times \mathbb{R}_{++}$ by letting $\ell := n+1$ and

$$x \; \mathsf{D}_{\Pi/\mathsf{L}} \; y \text{ if and only if } x_i/x_\ell \geq y_i/y_\ell \text{ for every } i = 1, \ldots, n, \text{ and } x/x_\ell \neq y/y_\ell \in \mathbb{R}^n. \tag{2.15}$$

Other examples of dominance orders are $\mathsf{D}_\Sigma$ defined in (4.15) and $\mathsf{D}_w$ defined in (4.20,4.21).

**Definition 2.14.** If there exist $x, y \in X$ such that $x \precsim y$ but $x \; \mathsf{D} \; y$, then we say that the market system admits D-*arbitrage*. If no such $(x,y)$ exist, then we say that the market system is D-*arbitrage-free*. For brevity the D may be suppressed.

Thus an arbitrageable market system allows some sequence of transactions to transform at least one state $x$ into a strictly worse state $y$. In Example 2.2 (respectively, Example 2.5), existence of $\mathsf{D}_\Pi$-arbitrage (resp., $\mathsf{D}_{\Pi/\mathsf{L}}$-arbitrage) would mean that from some starting state, there exists some sequence of swaps (resp., swaps and/or liquidity operations and/or gifts) that bleeds the reserves, in the sense of resulting in a net decrease in at least one of the asset balances (resp., asset balances per LP share), with no net increase in any of those balances (resp., balances per LP share).

On the surface, the concept of arbitrage in Definition 2.14 may seem to include only "internal" exploits within $M$, but that is not truly a restriction, because $M$ can be defined expansively to incorporate any relevant "external" oracles/venues/services, as in Example 4.14.

**Definition 2.15.** If a function $K : X \to \mathbb{R}$ satisfies

$$K(x) > K(y) \quad \text{for all } x, y \in X \text{ such that } x \; \mathsf{D} \; y, \tag{2.16}$$

then we say that $K$ is strictly D-*increasing*, which may be abbreviated as *increasing*.

The strictly D-*increasing* property, defined in (2.16) as a monotonicity with respect to the dominance order D on $X$, is distinct from, and not to be confused with, the *invariant* property, defined in (2.8) to entail monotonicity with respect to the $M$-induced preorder $\precsim$ on $X$. Thus an *increasing invariant* is a function that satisfies both the increasing (2.16) property and the invariant (2.8) property; but the word *increasing*, by itself, does not refer to the invariant (2.8) property.

**Definition 2.16.** If market system $M$ has an invariant, then recall Definition 2.8: $M$ is a *CFMM*.

If market system $M$ has a strictly D-*increasing* invariant, then we say that $M$ is a D-*CFMM*.

*MM* may stand for not just Market Maker, but also Market Mechanism or Market systeM; the last alternative emphasizes this framework's capability to model an entire network/ecosystem. The *CF* may stand for not just Constant Function (maintaining constant value on reversible operations) but also *Consistent* Function, or *Coherent* Function, referencing the consistency/coherence of the invariant with respect to the D dominance relation.

## 2.3 REMMs

This subsection gives several equivalent characterizations of a particular subcase of CFMMs, namely REMMs, which are a restrictive subcase, as we will discuss below.

**Definition 2.17.** A market system $M$ is said to be a *REMM* or *REversible* market system if

$$x \to y \text{ implies } y \to x. \tag{2.17}$$

Equivalently, a market system is a REMM if its induced preorder $\precsim$ is an equivalence relation. Equivalently, $M$ is a REMM if $\precsim_M$ coincides with $\sim_M$.

A special case of a general market system, a REMM by definition has a *symmetric* reachability relation $\precsim$, which is therefore an equivalence relation, which therefore partitions $X$ into *equivalence classes*. The reachable set from any state $x$ is exactly its equivalence class, denoted by $[x]$.

Outside of this subsection, our theorems will not assume the REMM condition (and will not make assumptions implying the REMM condition), because restricting to REMMs would exclude fee-collecting AMMs (and would also exclude donation-accepting AMMs) of importance in DeFi, such as uniswap.org v2 and curve.fi. Some of our examples are REMMs, but the theory is general.

**Definition 2.18.** Let $M$ be a market system. If there exists $K : X \to \mathbb{R}$ such that for all $x, y \in X$,

$$x \precsim_M y \text{ if and only if } K(x) = K(y), \tag{2.18}$$

then we say that $M$ is a €FMM, and that $K$ is a *€-invariant*, where the € symbol, pronounced "sequel", overlays C and = to express that (2.18) requires an *equal* invariant on all transitions.

All €FMMs are CFMMs, as (2.9) clearly holds, and (2.8) too since (2.18) rules out $x \prec y$.
All €-invariants are invariants, for the same reason.
The €FMMs are a narrow slice of all CFMMs, because all €FMMs are REMMs:

**Proposition 2.19.** *Let $X$ be a state space with cardinality not greater than continuum: $|X| \le |\mathbb{R}^n|$. A market system on $X$ is a €FMM if and only if it is a REMM.*

*Proof.* In any €FMM, if $x \to y$ then $K(x) = K(y)$ hence $y \to x$, which is the REMM property. Conversely, given a REMM, let $X/\sim$ denote the set of reachability-induced equivalence classes, and let $f$ be an injective function $X/\sim \to \mathbb{R}$, which exists since $|X| \le |\mathbb{R}^n| = |\mathbb{R}|$. Then $K(x) := f([x])$, where $[x]$ denotes the equivalence class of $x$, is a €-invariant, hence the REMM is a €FMM. □

Arbitrage-free REMMs have a simple characterization:

**Proposition 2.20.** *A REMM with €-invariant $K$ is arbitrage-free if and only if each equivalence class $S$ (equivalently: each level set $S$ of $K$) has no comparable states (no $x, y \in S$ with $x \mathsf{D} y$).*

*Proof.* Arbitrage ($x \precsim y$, $x \mathsf{D} y$) exists iff an equivalence class $S$ exists such that $x, y \in S$, $x \mathsf{D} y$. □

**Remark 2.21.** We comment on conditions assumed – or not assumed – in this paper.

First, the *no comparable states* condition in Proposition 2.20 is not assumed in Proposition 2.19. A similar condition (named NoDominatedTrades) *is* assumed in [14], together with other conditions that are collectively stronger than REMM (2.17), to form a joint condition that is shown, in that paper, to be sufficient for the €FMM[6] property. From the perspective of Proposition 2.19, REMM by itself (without any no-comparison/no-domination condition) is sufficient and necessary for the €FMM property.

Second, the REMM condition entails a restrictively strong form of path independence, in the following sense: For any sequence of states $\{x_i\}$ where $x_0 \to x_1 \to x_2 \to \cdots$, the reachable set from *any* $x_i$ is the *same* set, namely the equivalence class $S := [x_0] = [x_1] = [x_2] = \cdots$, with no "leakage" of attainable states. The reachable set from $x_i$ is independent not only of the path prior to $x_i$, but indeed independent of $x_i$ itself: it is $S$ for all $x_i \in S$. Our assumptions (outside of this subsection) do *not* entail this restrictively strong form of path independence that excludes fee-taking AMMs and gift-accepting AMMs.

Third, to be clear about "$x_0$", none of this paper's definitions assume a fixed initial state. In particular, *arbitrage-free* rules out arbitrage starting from *any* state in $X$, and invariants satisfy (2.8,2.9) for *all* pairs of states in $X$, not just states reachable from some fixed starting state. If there is a desire to fix an initial $x_0$, this framework is still applicable, by letting $X := M_\infty(x_0)$.

Fourth, let us juxtapose the reversibility (REMM) condition and the completeness condition. Reversibility holds if for all $x, y \in X$, we have either 0 or 2 of the relations $x \to y$ and $y \to x$. Completeness holds if for all $x, y \in X$, we have either 1 or 2 of those relations. Typical systems with *fees* are neither reversible (due to the existence of $x \prec y$) nor complete (due to existence of $x, y$ unreachable from each other). Relative to precursors within the €FMM (equivalently: REMM) special case, this paper (outside of this subsection) accommodates *general* market systems, not necessarily reversible or complete.

Fifth, our definition of *CFMM*, in agreement with the common-language usage of that term, includes fee-incorporating market makers such as Uniswap v2 and Curve. An alternative notion of CFMM that imposes the equality condition (2.18) would disagree with common terminology, as (2.18) excludes fee-incorporating protocols such as Uniswap v2 and Curve, which are commonly described as CFMMs. To disambiguate these conditions, we use the nomenclature €FMM for the subset of CFMMs that satisfy the equality-only condition (2.18), and we follow common terminology in defining CFMM broadly enough to allow pools that receive fees and gifts.

---

[6]in our terminology. The referenced paper does not have terminology distinguishing a €FMM from a general CFMM, as it does not investigate general CFMMs.

# 3 Necessary and sufficient conditions for absence of arbitrage

## 3.1 Sufficient conditions for absence of arbitrage

The fact that certain CFMMs (including certain REMMs, and modified versions that take fees) rule out arbitrage goes back at least as far as the early-stage discussions among DeFi pioneers, for instance Buterin [7]:

> What this means is that a "reserve bleeding" attack on a market maker that preserves this kind of path independence property is impossible. Even if some trolls successfully create a market panic that drops prices to near-zero, when the panic subsides, and prices return to their original levels, the market maker's position will be unchanged – even if both the price, and the market maker's balances, made a bunch of crazy moves in the meantime...
>
> we certainly can modify the market maker to earn revenue, and quite simply: we have it charge a spread ... This sacrifices the "path independence" property, but in such a way that any deviations from path independence are always in the market maker's favor.

Indeed this is universally true, in the following sense:

**Theorem 3.1.** *Every* D*-CFMM is* D*-arbitrage-free.*

*Proof.* For any $x, y \in X$ with $x \precsim y$, we have $K(x) \le K(y)$ by Definition 2.8. Because $K$ is strictly D-increasing, it cannot be true that $x$ D $y$. □

Theorem 3.1 makes clear the universality of the thesis advanced by Buterin [7]. It assumes only the monotonicity of the invariant, not continuity (as in [6]) of the invariant[7] nor any other assumption. It is valid irrespective of how many moves are referenced by the *bunch* of moves. It remains valid irrespective of how we define *bleeding*, as long as D embodies that notion of bleeding; in particular, it is still valid if *balances* are replaced by balances per liquidity share. It even remains valid irrespective of what types of actions the trolls perform – swaps, or liquidity provision/removal, or anything else – provided that the invariant property holds for all operations.

## 3.2 Zeroth fundamental theorem: existence of invariants

Existence of invariants can be guaranteed by topological conditions. Endow $X$ with a topology $\mathcal{X}$ that is *second countable*, meaning that $\mathcal{X}$ has a countable base, where *countable* means finite or countably infinite. In the case of uncountable $X \subseteq \mathbb{R}^n$, an example of a second countable $\mathcal{X}$ is the standard Euclidean topology. In the case of countable $X$, it will be useful to let $\mathcal{X}$ be the discrete topology.

---

[7]Nowhere do we assume continuity of the *invariant*. (In the next sections, we assume a different sense of continuity, namely order-continuity of the market system, to formulate the *converse*, in Theorem 3.11.)

**Definition 3.2.** Define a market system to be *order-continuous* (or more briefly: *continuous*) if its induced strict preorder $\prec$ is continuous, meaning that the strict transition sets to $x$ and from $x$

$$\begin{aligned} L_M(x) &:= \{y \in X : y \prec_M x\}, \\ G_M(x) &:= \{y \in X : x \prec_M y\} \end{aligned} \quad (3.1)$$

are open (with respect to $\mathcal{X}$) for all $x \in X$.

For example, every REMM $M$ is continuous, because $L_M = G_M = \emptyset$.

Market systems which are not continuous may still extend to continuous market systems.

**Definition 3.3.** Consider market systems $R$ and $E$ on $X$. We say that $R$ *extends to* $E$ (or that $R$ is a *restriction of* $E$, or that $E$ is an *extension of* $R$) if the induced preorders $\precsim_R$ and $\precsim_E$ satisfy

$$x \precsim_E y \text{ for all } x, y \in X \text{ such that } x \precsim_R y, \quad (3.2)$$

and the induced strict partial orders $\prec_R$ and $\prec_E$ satisfy

$$x \prec_E y \text{ for all } x, y \in X \text{ such that } x \prec_R y. \quad (3.3)$$

In other words, the relation $\precsim_R$ is a subset of $\precsim_E$, and the relation $\prec_R$ is a subset of $\prec_E$.

**Lemma 3.4.** *If $R$ extends to $E$, then any invariant for $E$ is an invariant for the restriction $R$.*

*Proof.* Applying (3.2) to $(x, y)$ and to $(y, x)$ shows that $x \sim_E y$ for all $x, y \in X$ such that $x \sim_R y$, So if $K$ satisfies (2.9) for $\sim_E$ then it does for $\sim_R$. The analogous statement (2.8) for $\prec_E$ and $\prec_R$ is by (3.3). $\square$

**Definition 3.5.** An *AMM* is an order-continuous market system, or a restriction thereof.

This definition is natural because continuity always holds in a truly *automated* system; the idea of automation implies a machine with a countable state space, and thereby the existence of a second-countable $\mathcal{X}$, namely the discrete topology, that automatically implies *continuity*.

All AMMs are CFMMs. To prove this, we apply Sondermann's generalization, to *partial* orders, of Debreu's [11] utility representation of total orders. Continuity rules out situations such as 4.23.

**Theorem 3.6** (Zeroth fundamental theorem). *All AMMs are CFMMs.*

*In other words, if $M$ is, or extends to, an order-continuous market system, then $M$ is a CFMM; and moreover $M$ has both an upper-semicontinuous and a lower-semicontinuous invariant.*

*Proof.* Let $E$ be a continuous market system extending $M$. Applying Sondermann [18] Theorem 2 to the strict partial order induced by $E$, there exists an upper-semicontinuous $K$, and also a lower-semicontinuous $K$, satisfying (2.8) and indeed a stronger conclusion than (2.9), namely that $K(x) = K(y)$ for all $x, y$ such that $L_E(x) = L_E(y)$ and $G_E(x) = G_E(y)$. Thus $K$ is invariant for $E$, and therefore also for $M$. $\square$

Order-continuity always holds in practice, by Corollary 3.7.

**Corollary 3.7.** *Every market system on a countable space is continuous and is therefore a CFMM.*

*Proof.* Let $M: X \to 2^X$ be a market system on a countable state space $X$. Let $\mathcal{X}$ be the discrete topology, which is therefore second-countable, and which makes $M$ continuous because all sets are open. Then $M$ has an invariant, by Theorem 3.6. □

Even in the uncountable case, there are simple sufficient conditions for a market system to be a restriction of a continuous market system, namely that the 1-step transition sets be subsets of open sets that are nested in the following sense.

**Proposition 3.8** (Sufficient condition for continuity in the uncountable case). *If $M$ and $M^*$ are market systems such that, for all $x \in X$, we have $M(x) \subseteq M^*(x)$ and*

(a) *The 1-transition sets of $M^*$ from/to $x$, namely $M^*(x)$ and $\{v : x \in M^*(v)\}$, are open, and*

(b) *The 1-transition sets of $M^*$ are nested: $x \notin M^*(y) \subseteq M^*(x)$ for all $y \in M^*(x)$.*

*Then $M^*$ is continuous, and $M$ extends to $M^*$*

*Proof.* For all $x \in X$ we have $M^*_\infty(x) = \{x\} \cup M^*(x)$, and for all $y \in M^*(x)$ we have $M^*_\infty(y) \subseteq M^*(x)$ which does not contain $x$. Thus $\{y : x \prec y\} = M^*_\infty(x) \setminus \{x\} = M^*(x)$ which is open. Likewise $\{v : v \prec x\} = \{v : x \in M^*_\infty(v)\} \setminus \{v : v \in M^*_\infty(x)\} = \{x\} \cup \{v : x \in M^*(v)\} \setminus \{x\}$ is open. Thus $M$ is continuous. It remains to show that $M$ extends to $M^*$. From $M(x) \subset M^*(x)$ we have $\precsim$ implies $\precsim^*$. To show that $\prec$ implies $\prec^*$, we have, for all $x \prec y$, that $x \neq y \in M^*_\infty(x)$, hence $y \in M^*(x)$, hence $x \notin M^*_\infty(y)$, as claimed. □

Thus the continuity condition in Theorem 3.6 / Definition 3.10 does not require checking (3.1); it is enough that $M(x) \subseteq M^*(x)$ where $M^*$ has nested open 1-step transition sets.

### 3.3 Necessary conditions for absence of arbitrage

Theorem 3.11 will show, assuming continuity and giftability, that any *arbitrage-free* market system must have an increasing invariant.

**Definition 3.9.** A market system on $X$ is *giftable* if $x \precsim y$ for all $x, y \in X$ satisfying $y \mathrel{\mathsf{D}} x$.

We use the terms *gift* and *donation* synonymously. The primary intuition is that giftability holds if the system's asset wallets "accept donations"; but even for systems that do not take gifts, a secondary intuition is that giftability automatically holds, if the space $X$ is defined narrowly enough, that it does not contain any pairs $x, y$ with $y \mathrel{\mathsf{D}} x$.

**Definition 3.10.** A *DeFi market system* (or *DeFi market* or *DeFi system*) is a giftable market system that is, or extends to, a continuous market system.

This definition makes sense, given DeFi's home on the blockchain. Every market system implemented by a smart contract on a blockchain has a countable state space and is therefore *continuous* by Corollary 3.7; moreover, blockchain wallets which store assets of DeFi protocols ordinarily accept gifts (or, to use the terminology from [1], *donations*).

For DeFi systems, this converse to Theorem 3.1 gives a necessary condition for no-arbitrage:

**Theorem 3.11.** *If $M$ is a $\mathsf{D}$-arbitrage-free DeFi system, then $M$ is a $\mathsf{D}$-CFMM.*

*Proof.* By Theorem 3.6, the continuous extension has an invariant $K$, which is also an invariant for $M$. For any $x, y \in X$ with $y \mathrel{\mathsf{D}} x$, we have $x \precsim y$ by giftability. By absence of arbitrage, $x \prec y$, therefore $K(x) < K(y)$. Thus $K$ is strictly increasing, and $M$ is a $\mathsf{D}$-CFMM. □

### 3.4 First fundamental theorem

In linear[3] markets, the first Fundamental Theorem of Asset Pricing establishes the equivalence of no-arbitrage[4] and existence of a pricing measure. In nonlinear DeFi markets, the parallel result establishes the equivalence of no-arbitrage and the existence of an increasing invariant.

**Theorem 3.12.** *A DeFi system is arbitrage-free if and only if it has an increasing invariant.*

*Proof.* Combine Theorems 3.1 and 3.11. □

The arbitrage, giftability, and monotonicity properties are all relative to the same dominance relation. That notion of dominance has full flexibility: *any* strict partial order – or *orders* – on $X$ may be chosen; a CFMM can be proven to be $\mathsf{D}_1$-arbitrage free because it has a $\mathsf{D}_1$-increasing invariant, and moreover proven to be $\mathsf{D}_2$-arbitrage free because it has a $\mathsf{D}_2$-increasing invariant.

## 4 Applications

As an immediate application of the "only if" direction of the first fundamental theorem, we have:

**Remark 4.1.** For the purpose of designing an arbitrage-free DeFi system, Theorem 3.12 shows that the designer does not need to consider general market systems. The design space can be restricted to include only CFMMs with increasing invariants, because those are precisely the arbitrage-free systems. This remark applies not just to DEX design, but also to *any* DeFi protocol design.

The following applications of the "if" direction of the first fundamental theorem are facilitated by a lemma, that a function which satisfies a "one-step" version of invariance is in fact an invariant:

**Lemma 4.2.** *Let $M$ be a market system on $X$. If there exists $K : X \to \mathbb{R}$ satisfying*

$$K(x) < K(y) \quad \text{for all } x, y \in X \text{ such that } y \in M(x) \text{ and } x \notin M(y), \tag{4.1}$$

$$K(x) = K(y) \quad \text{for all } x, y \in X \text{ such that } y \in M(x) \text{ and } x \in M(y), \tag{4.2}$$

*then $M$ has invariant $K$.*

*Proof.* For all $z \in M_\infty(x)$ there exists a sequence $x = z^0, z^1, \ldots, z^k = z$, where each $z^{i+1} \in M(z^i)$, which proves $K(x) \leq K(z)$. If $K(x) = K(z)$ holds, then each $K(z^i) = K(z^{i+1})$, hence $z^i \sim z^{i+1}$, which proves $x \sim z$ and verifies the invariant property. □

Proposition 4.3 shows that a typical mechanism collecting a fee $\phi(x,y) \in \mathbb{R}^n$ for a transaction from state $x$ to state $y$ admits an increasing invariant and hence does not allow arbitrage:

**Proposition 4.3** (Typical fee mechanisms admit invariants)**.** *Let $M$ be a market system on some $X \subseteq \mathbb{R}^n$. Consider a fee structure $\phi : X \times X \to \mathbb{R}^n$ such that $\phi(x,x) = 0$ and $y$ D $(y - \phi(x,y))$ for all $x \neq y$, where D is a dominance relation on $\Phi := X \cup \{y - \phi(x,y) : (x,y) \in X^2\}$.*

*If there exists a strictly D-increasing $\kappa : \Phi \to \mathbb{R}$ such that, for all $x \in X$*

$$M(x) \subseteq \{y \in X : \kappa(y - \phi(x,y)) \geq \kappa(x)\}, \tag{4.3}$$

*then $M$ is a D-CFMM with invariant $\kappa$ (restricted to $X$) and is therefore D-arbitrage free.*

*Proof.* For all $x \in X$, all $y \in M(x)$, we have $\kappa(y) \geq \kappa(y - \phi(x,y)) \geq \kappa(x)$ with equality if $x = y$, and only if $x = y$, because otherwise $\kappa(y) > \kappa(y - \phi(x,y))$. This suffices, by Lemma 4.2. □

AMM whitepapers routinely display increasing invariants as a function of the asset balances, but not the liquidity supply. Proposition 4.4 implies that an increasing invariant on asset balances induces an increasing invariant on the space augmented to include liquidity provision/removal. Thus the guarantee that swap operations are unexploitable extends to a guarantee that combinations of swap and liquidity operations are unexploitable, provided that the transaction amounts are properly determined, as done in (4.5) or (4.4). Example 4.15 shows a type of improper design.

For $x \in \mathbb{R}^n$ and $\xi \in \mathbb{R}$, let $x \oplus \xi$ denote the concatenation $(x_1, x_2, \ldots, x_n, \xi) \in \mathbb{R}^{n+1}$.

**Proposition 4.4** (Properly introducing liquidity operations does not introduce arbitrage)**.** *Let $X \subseteq \mathbb{R}^n_+$ and $\ell := n+1$. Let $\kappa : X \to \mathbb{R}$ be strictly $\mathsf{D}_\sqcap$-increasing. For some $\xi > 0$, let $\bar{M}$ be a market system on $\bar{X} := \{\alpha x \oplus \alpha \xi : \alpha > 0, x \in X\} \subseteq \mathbb{R}^\ell_+$. Assume that either (i) For all $x \oplus x_\ell \in \bar{X}$,*

$$\bar{M}(x \oplus x_\ell) \subseteq \{\alpha x \oplus \alpha x_\ell : \alpha > 0\} \cup \Big\{y \oplus y_\ell \in \bar{X} : \kappa\Big(\frac{\xi}{y_\ell}y\Big) > \kappa\Big(\frac{\xi}{x_\ell}x\Big)\Big\}, \tag{4.4}$$

*or (ii) $\kappa$ is an invariant for a market system $M$ on $X$, and for all $x \oplus x_\ell \in \bar{X}$,*

$$\bar{M}(x \oplus x_\ell) = \{\alpha x \oplus \alpha x_\ell : \alpha > 0\} \cup \Big\{y \oplus y_\ell \in \bar{X} : y_\ell = x_\ell \text{ and } \frac{\xi}{y_\ell}y \in M\Big(\frac{\xi}{x_\ell}x\Big)\Big\}. \tag{4.5}$$

*Then $\bar{M}$ is a $\mathsf{D}_{\sqcap/\mathsf{L}}$-CFMM and is therefore $\mathsf{D}_{\sqcap/\mathsf{L}}$-arbitrage free.*

Both versions (4.4,4.5) allow liquidity operations $x \oplus x_\ell \to \alpha x \oplus \alpha x_\ell$, where $\alpha < 1$ for removal and $\alpha > 1$ for addition. The versions differ in that (4.5) builds on $M$ without introducing any further fees (beyond whatever fees were already in $M$), whereas version (4.4) builds on a function $\kappa$ (without referencing any $M$) and allows gifts and fees (including fees collected *from* the pool by dilution, as in Example 2.6).

Proposition 4.4 does not assume that the underlying $X$ is a cone, nor that $\kappa$ is homogeneous.

*Proof.* By applying Lemma 4.2, let us show that $\bar{M}$ has invariant $K : \bar{X} \to \mathbb{R}$ where

$$K(x \oplus x_\ell) := \kappa\Big(\frac{\xi}{x_\ell}x\Big). \tag{4.6}$$

In either case (4.4,4.5), for any $x \oplus x_\ell \in \bar{X}$, any $y \oplus y_\ell \in \bar{M}(x \oplus x_\ell)$, we have

$$K(x \oplus x_\ell) = \kappa\Big(\frac{\xi}{x_\ell}x\Big) \le \kappa\Big(\frac{\xi}{y_\ell}y\Big) = K(y \oplus y_\ell). \tag{4.7}$$

In case (4.4) if equality holds in (4.7), then $y/y_\ell = x/x_\ell$, so $(x \oplus x_\ell) \sim_{\bar{M}} (y \oplus y_\ell)$ and we are done.

In case (4.5), either $y/y_\ell = x/x_\ell$ hence we are again done, or else $\xi y/y_\ell \in M(\xi x/x_\ell)$ so if equality holds in (4.7) then $\xi x/x_\ell \in M(\xi y/y_\ell)$ because $\kappa$ is invariant for $M$, so $(x \oplus x_\ell) \sim_{\bar{M}} (y \oplus y_\ell)$ as desired. Therefore $K$ is invariant. It is $\mathsf{D}_{\Pi/\mathsf{L}}$-increasing because $(x \oplus x_\ell)\mathsf{D}_{\Pi/\mathsf{L}}(y \oplus y_\ell)$ implies $(\xi x/x_\ell)\mathsf{D}_\Pi(\xi y/y_\ell)$ hence $K(x \oplus x_\ell) > K(y \oplus y_\ell)$. □

Therefore allowing liquidity provision/removal operations in combination with swaps does not introduce any exploits that could leave the AMM with worse balances per liquidity share.

## 4.1 AMMs

**Example 4.5** (Uniswap v2 with LP fees and LP operations, stylized). The market system $\bar{M} := M'$ from Example 2.5 satisfies (4.4) with $\kappa(x_1, x_2) := x_1 x_2$ and $\xi = 1$, so by Proposition 4.4, it is a $\mathsf{D}_{\Pi/\mathsf{L}}$-CFMM with increasing invariant

$$K(x_1, x_2, x_\ell) = x_1 x_2 / x_\ell^2. \tag{4.8}$$

Therefore Example 2.5, a stylized Uniswap v2 with fees and LP operations, is $\mathsf{D}_{\Pi/\mathsf{L}}$-arbitrage free.

**Example 4.6** (LP fees and admin fees). Let $\bar{M} := M''$ from Example 2.6 in the style of Sushiswap, which forked – but activated – the Uniswap admin fee collection (of one-sixth of the 0.3% fee). Then $\bar{M}$ satisfies (4.4) with $\kappa(x) = \kappa(x_1, x_2) := \sqrt{x_1 x_2}$ and $\xi = 1$, because for all $y \oplus y_\ell \in \bar{M}(x \oplus x_\ell)$, by (2.4), either $y \oplus y_\ell = \alpha(x \oplus x_\ell)$ for some $\alpha > 0$, or

$$\frac{\kappa(y)}{y_\ell} = \frac{\kappa(y)}{x_\ell} \Big/ \frac{6\kappa(y)}{5\kappa(y) + \kappa(x)} = \frac{1}{x_\ell}\Big(\frac{5}{6}\kappa(y) + \frac{1}{6}\kappa(x)\Big) > \frac{\kappa(x)}{x_\ell}. \tag{4.9}$$

By Proposition 4.4, therefore, Example 2.6 has $\mathsf{D}_{\Pi/\mathsf{L}}$-increasing invariant $K(x_1, x_2, x_\ell) = \sqrt{x_1 x_2}/x_\ell$, and cannot be $\mathsf{D}_{\Pi/\mathsf{L}}$-exploited by any sequence of swaps and/or liquidity operations.

**Example 4.7** (Curve StableSwap, stylized). Let $n \ge 2$ and $X := \mathbb{R}_{++}^n$. Each state $x \in X$ consists of the balances of $n$ assets in the pool. Let $A > n^{-n}$. Following [12], define $f : \mathbb{R}_+ \times X \to \mathbb{R}$ by

$$f(D, x) := An^n \sum x_i + D - ADn^n - \frac{D^{n+1}}{n^n \prod x_i}. \tag{4.10}$$

Then $f(0, x) > 0$ and $\frac{\partial f}{\partial D}(D, x) < 1 - An^n < 0$ for all $x \in X$, all $D > 0$. So for each $x \in X$, define $\kappa(x)$ to be the unique root of $f(\cdot, x)$.

The stylized (no-fee, no-LP-operations, no-roundoff) Curve StableSwap AMM

$$M(x) := \{y \in X : \kappa(y) \geq \kappa(x)\} \tag{4.11}$$

clearly has invariant $\kappa$. By Proposition 4.8 the invariant is $\mathsf{D}_\Pi$-increasing, and the stylized Curve StableSwap is therefore $\mathsf{D}_\Pi$-arbitrage free.

**Proposition 4.8.** *The Curve Stableswap invariant $\kappa$ is $\mathsf{D}_\Pi$-increasing and homogeneous.*

*Proof.* We have $f(\alpha D, \alpha x) = \alpha f(D, x)$ for all $\alpha > 0$, all $x \in X$, which implies

$$f(\alpha\kappa(x), \alpha x) = \alpha f(\kappa(x), x) = 0, \text{ hence } \kappa(\alpha x) = \alpha \kappa(x), \tag{4.12}$$

proving homogeneity of the invariant. To show $\kappa$ increasing, consider $x, y \in X$ with $x\ \mathsf{D}_\Pi\ y$. Then

$$f(\kappa(y), y) = 0 = f(\kappa(x), x) < f(\kappa(x), y), \tag{4.13}$$

but $f(\cdot, y)$ is strictly decreasing, so $\kappa(y) > \kappa(x)$ as claimed. □

## 4.2 Systems of AMMs

The "system" of AMMs given below, by $M$ in (4.14) or by $M$ in (4.18), could simply be called an AMM by our definition, but adding "system" helps convey the compound nature of each example.

**Example 4.9** (Independent AMMs on the same assets)**.** Define $M$ and dominance order $\mathsf{D}_\Sigma$ by

$$M(x_1, x_2, x_3, x_4) := \{(y_1, y_2, y_3, y_4) \in X : y_1 y_2 = x_1 x_2 \text{ and } y_3 y_4 = x_3 x_4\}, \quad X := \mathbb{R}^4_{++}, \tag{4.14}$$

$$x\ \mathsf{D}_\Sigma\ y \text{ if } x_1 + x_3 \geq y_1 + y_3 \text{ and } x_2 + x_4 \geq y_2 + y_4, \text{ with at least one inequality strict.} \tag{4.15}$$

This models a system of two constant-product AMMs, where $(x_1, x_2)$ are the balances of two assets in the first AMM, and $(x_3, x_4)$ are the balances of those *same* two assets in the second AMM. Then the DeFi system $M$ admits $\mathsf{D}_\Sigma$-arbitrage; indeed $M$ allows any state with $x_1/x_2 \neq x_3/x_4$ to be transformed to a $\mathsf{D}_\Sigma$-worse state. For example $(8, 2, 1, 9) \to (4, 4, 3, 3)$ but $(8, 2, 1, 9)\mathsf{D}_\Sigma(4, 4, 3, 3)$.

Neither of the two AMMs admits arbitrage *individually*; to be precise, $M$ admits neither $\mathsf{D}_{1,2}$-arbitrage nor $\mathsf{D}_{3,4}$-arbitrage, where $x\ \mathsf{D}_{i,j}\ y$ is defined to mean that $x_i \geq y_i$ and $x_j \geq y_j$, with at least one inequality strict. However, the two AMMs *collectively* admit arbitrage, as the AMM *system* allows *aggregate* holdings of $(8+1, 2+9) = (9, 11)$ to be reduced to a $\mathsf{D}_\Sigma$-worse state, with aggregate holdings of $(7, 7)$.

Arbitrage in Example 4.9 arises from availability of different prices (exchange rates) in different venues (liquidity pools, in this case) of the market system. Complementary to research [5, 9] on maximizing an arbitrageur's profits in certain swaps-only DEX setups, this paper's Theorem 3.12 characterizes (without optimizing) the existence/absence of generalized arbitrage in *general* DeFi systems, including non-DEX services, and DEXes that allow both liquidity operations and swaps.

A system that has multiple pools can avoid arbitrage, if those pools are sufficiently coordinated:

**Example 4.10** (Linked AMMs: the case of Uniswap v3). Let $0 < r_0 < r_1 < \cdots < r_J$, which can be described as the tick marks, quoted as square roots of prices, that separate the price ranges.

Define the state space

$$X = \{(L, R) \in \mathbb{R}_+^J \times [r_0, r_J] : L \neq 0 \in \mathbb{R}^J\}. \tag{4.16}$$

Following [2], define the functions $x_j, y_j : X \to \mathbb{R}_+$ by

$$\begin{aligned} x_j(L, R) &:= L_j \Big(\frac{1}{C_j(R)} - \frac{1}{r_j}\Big), \\ y_j(L, R) &:= L_j \Big(C_j(R) - r_{j-1}\Big), \end{aligned} \tag{4.17}$$

where $C_j(r) := r_{j-1} \vee r \wedge r_j$ is the clamp function $\mathbb{R} \to [r_{j-1}, r_j]$ onto the $j$th price range.

The $j$th pool has balances $(x_j, y_j)$ of (asset 1, asset 2). The $L_j$ can be described as the amount of liquidity in the $j$th pool, and $R^2$ can be described as the price of asset 1 in units of asset 2. Let $\Lambda(L) := \{j \in \{1, \ldots, J\} : L_j > 0\}$ denote the set of nonempty pools. On $X$ define the AMM

$$M(L, R) = \{(L', R') \in X : \text{Either (i) } L' = L \text{ or (ii) } R = R'\}. \tag{4.18}$$

which is a stylized no-roundoff Uniswap v3. It supports (i) swaps between assets 1 and 2, that change $x$ and $y$ in some pool(s) but maintain ($L' = L$) the combined liquidity in each pool, and (ii) liquidity operations that change the combined liquidity in some pool(s) but maintain the asset proportions in each pool: changing $L_j$ to $L'_j$ requires scaling the asset reserves $x_j$ and $y_j$ by $L'_j/L_j$, as (4.17) shows. Unlike Uniswap v2, fees in v3 accumulate *outside* of the liquidity pools, so (4.17) does not track fees. Extra state variables could track fee accumulation, but that would not affect the $(L, R)$ specification; rather such fee variables could be referenced in defining a dominance order that accounts for fees accruing to LPs. This example will define dominance purely on pool balances.

Absence of arbitrage in the $j$th pool for each *individual* $j = 1, \ldots J$ can be proved by applying Proposition 4.4, but the larger question is whether the *system* of pools remains arbitrage-free in aggregate. In Example 4.9 the aggregated system failed to remain arbitrage-free. For the system in Example 4.10, consider the following "weighted" $\mathsf{D}_w$ generalizing $\mathsf{D}_\Sigma$ dominance:

For $w \in \mathbb{R}_+^J$ define the strict partial order $\mathsf{D}_w$ on $X$ by $(L, R)\mathsf{D}_w(L', R')$ if

$$\Lambda_w \subseteq \Lambda(L) \cap \Lambda(L') \qquad \text{where } \Lambda_w := \{j : w_j > 0\}, \tag{4.19}$$

and

$$\sum_{j \in \Lambda_w} \frac{w_j}{L_j} x_j(L, R) \geq \sum_{j \in \Lambda_w} \frac{w_j}{L'_j} x_j(L', R'), \tag{4.20}$$

and

$$\sum_{j \in \Lambda_w} \frac{w_j}{L_j} y_j(L, R) \geq \sum_{j \in \Lambda_w} \frac{w_j}{L'_j} y_j(L', R'), \tag{4.21}$$

and at least one of the two inequalities is strict.

This $\mathsf{D}_w$ notion of dominance is motivated by the perspective of a liquidity provider (LP) who has supplied liquidity $w_j \geq 0$ to the $j$th range (where each $w_j \leq L_j \wedge L'_j$, although the weaker condition (4.19) suffices for our calculations). The LP passively keeps $w$ unchanged, while potential attackers perform sequences of (i) swaps and/or (ii) liquidity operations that may change $L$ to $L'$. If the amounts of assets 1 and 2 owned by this LP (in aggregate, across all of his or her pools) can ever be exploited into a dominated state (with aggregate amounts that have not increased in either asset, and decreased in at least one asset), then arbitrage exists, from this LP's perspective. Accordingly, those aggregate amounts are calculated in (4.20, 4.21), and $\mathsf{D}_w$ expresses this dominance concept.

In contrast to Example 4.9, this system remains arbitrage-free in aggregate:

**Proposition 4.11.** *The stylized Uniswap v3 system* (4.18) *is $\mathsf{D}_w$-arbitrage free.*

*Proof.* For any state $(L, R) \in X$ we have

$$\sum_{j \in \Lambda_w} \frac{w_j}{L_j} x_j(L, R) = \sum_{j \in \Lambda_w} w_j \Big(\frac{1}{C_j(R)} - \frac{1}{r_j}\Big) \text{ and } \sum_{j \in \Lambda_w} \frac{w_j}{L_j} y_j(L, R) = \sum_{j \in \Lambda_w} w_j \Big(C_j(R) - r_{j-1}\Big) \quad (4.22)$$

and likewise for any $(L', R') \in X$. The right-hand sides are, respectively, decreasing and increasing in $R$. So if (4.20) then $R \leq R'$, and if (4.21) then $R \geq R'$; and if both (4.20, 4.21), then $R = R'$, in which case both (4.20, 4.21) are equalities. Hence $(L, R)\mathsf{D}_w(L', R')$ is impossible, and $\mathsf{D}_w$-arbitrage does not exist in the stylized Uniswap v3. $\square$

Unlike most AMM analyses, the state space (4.16) does not explicitly include asset balances. Instances of the general market system framework can employ whatever state variables are suitable in the application at hand – in this case, the liquidity-price variables $L \in \mathbb{R}_+^J$ and $R \in \mathbb{R}_+$ from [2].

### 4.3 Counterexamples

One direction of the first fundamental theorem was straightforward: that an increasing invariant rules out arbitrage. The converse, that all arbitrage-free DeFi systems have increasing invariants, required more care, for reasons seen in Examples 4.12 and 4.13, which show what can happen without continuity and giftability, respectively. In both examples, the state space $X \subseteq \mathbb{R}_+^2$ tracks balances of two assets, while arbitrage/increasing are defined by the dominance relation $\mathsf{D} = \mathsf{D}_\sqcap$.

**Example 4.12** (An arbitrage-free market system with no invariant)**.** Debreu's [11] example from utility theory adapts directly to DeFi theory. On the space $X := [0, 1]^2$, define the lexicographic market system

$$M(x_1, x_2) = \{(y_1, y_2) \in X : y_2 > x_2 \text{ or } (y_2 = x_2 \text{ and } y_1 > x_1)\}. \quad (4.23)$$

Indeed $M$ is an arbitrage-free market system: if $x \precsim y$ then $x_1 < y_1$ or $x_2 < y_2$, hence not $x \mathsf{D} y$. However, the existence of an invariant $K$ would imply the existence, for each $z \in [0, 1]$, of a rational $q_z \in (K(0, z), K(1, z))$, but then $\{q_z : z \in [0, 1]\}$ would be an uncountable set of rationals. Hence $M$ has no invariant.

**Example 4.13** (An arbitrage-free market system with an invariant but no increasing invariant)**.** Define the market system $M$ on $X := \{(1,3),(1,4),(2,1),(2,2)\}$ by

$$M(1,4) = \{(2,1)\}, \qquad M(2,2) = \{(1,3)\}, \qquad M(2,1) = M(1,3) = \emptyset. \tag{4.24}$$

Then $M$ is arbitrage-free (neither of the two possible trades is a D-arbitrage), and has an invariant, but it does not have a strictly D-increasing invariant; because any such invariant $K$ would satisfy

$$K(1,4) \leq K(2,1) < K(2,2) \leq K(1,3) < K(1,4), \tag{4.25}$$

which is impossible.

The next example shows that not all CFMMs admit homogeneous invariants, and the subsequent example shows what can go wrong in the nonhomogenous case, if interaction between swaps and liquidity operations is handled by a naive market design inconsistent with Proposition 4.4.

**Example 4.14** (Existence of an invariant does not imply the existence of a homogeneous invariant)**.** To show that the absence of a homogeneous invariant can occur with or without fees, we give two examples of AMMs: one with and one without fees. The no-fee case follows the classical example of nonhomothetic preferences. On $X = \mathbb{R}_{++}^2$ define the AMMs $M^{(0)}$ and $M^{(0.2)}$ where

$$M^{(c)}(x_1, x_2) = \{(y_1, y_2) \in X : y_1 - \phi_1 + \sqrt{y_2 - \phi_2} \geq x_1 + \sqrt{x_2}\} \text{ with } \phi_j := c\max(y_j - x_j, 0), \tag{4.26}$$

and the constant parameter $c := 0$ in the no-fee case, or $c := 0.2$ in a 20%-fee case. In either case, $M^{(c)}$ has an invariant, for example the function $(x_1, x_2) \mapsto x_1 + \sqrt{x_2}$. However $M^{(0.2)}$ does not have any invariant $K$ that is homogeneous (meaning $K(\alpha x) = \alpha K(x)$ for all $x \in X$, $\alpha > 0$), because $(4,16) \prec_{M^{(0.2)}} (6.8, 4)$ implies that a hypothetical homogeneous invariant for $M^{(0.2)}$ would satisfy

$$4K(1,4) = K(4,16) < K(6.8, 4) = 4K(1.7, 1), \tag{4.27}$$

which contradicts $(1.7, 1) \prec_{M^{(0.2)}} (1, 4)$. The same argument (using the same numerical states) is valid also for $M^{(0)}$, which shows that $M^{(0)}$ also does not have any invariant that is homogeneous.

**Example 4.15** (Arbitrage due to naive introduction of liquidity operations)**.** Suppose that liquidity shares were to be introduced into (4.26) by what we call the *naive* design: On $\bar{X} := \mathbb{R}_{++}^3$ define

$$\bar{M}^{(c)}(x_1, x_2, x_\ell) := \{\alpha(x_1, x_2, x_\ell) : \alpha > 0\} \cup \{(y_1, y_2, y_\ell) : y_\ell = x_\ell \text{ and } (y_1, y_2) \in M^{(c)}(x_1, x_2)\},$$

which ignores the (4.5) swap rule $(\xi/y_\ell)y \in M^{(c)}((\xi/x_\ell)x)$, and instead naively uses $y \in M^{(c)}(x)$. This $\bar{M}^{(c)}$ admits $\mathsf{D}_{\Pi/\mathsf{L}}$-arbitrage $(1.8, 1, 1) \to (1, 4, 1) \to (4, 16, 4) \to (6.8, 4, 4) \to (1.7, 1, 1)$ in both cases $c = 0$ and $c = 0.2$. This naive approach would be arbitrage-free if a homogeneous invariant on $X$ could recover $M^{(c)}$, but that is not the case in (4.26), where no homogeneous invariant exists. Proposition 4.4 handles liquidity operations robustly, including the nonhomogenous case.

# 5 A CFMM has a unique invariant if and only if it is complete

In linear[3] markets, the second Fundamental Theorem of Asset Pricing shows the equivalence of market completeness[5] and the uniqueness of the pricing measure. The DeFi analogue is as follows.

**Theorem 5.1** (Second fundamental theorem). *A CFMM (a market system with invariant) has a unique invariant if and only if it is complete.*

*Here "uniqueness" means modulo strictly increasing transformation – specifically, uniqueness with respect to the equivalence relation (on the set of functions $X \to \mathbb{R}$) defined by: $K_1$ and $K_2$ are equivalent if there exists a strictly increasing $f : K_2(X) \to \mathbb{R}$ such that $K_1 = f \circ K_2$.*

*Proof of "only if".* Consider a market system with unique invariant $K$. To show it is complete, suppose that there exist $u, v \in X$ such that neither $u \precsim v$ nor $v \precsim u$. Relabeling $u$ and $v$ if necessary, we have $K(u) \geq K(v)$. Let $c := K(u) - K(v) + 1$.

Define $K^* : X \to \mathbb{R}$ by $K^*(x) := K(x)$ if $x \precsim u$, otherwise $K^*(x) := K(x) + c$.

To show that $K^*$ is also an invariant, we will show, for any $x, y \in X$, that $K^*(y) - K^*(x)$ has the correct sign. If the conditions $x \precsim u$ and $y \precsim u$ are both true or both false, then $K^*(y) - K^*(x) = K(y) - K(x)$, as desired. If exactly one of those two conditions is true, let us say $x \precsim u$, then it cannot be true that $y \precsim x$. So if not $x \precsim y$, then nothing needs to be verified about $K^*(y) - K^*(x)$; on the other hand, if $x \precsim y$, then $x \prec y$, therefore $K^*(x) = K(x) < K(y) < K^*(y)$ which completes the proof that $K^*$ is invariant.

Moreover $K^*(u) - K^*(v) < 0 \leq K(u) - K(v)$ implies that $K^*$ is not a strictly increasing transformation of $K$, contradicting uniqueness. Therefore no such $u, v$ exist. □

*Proof of "if".* Suppose a complete market system has invariants $K_1$ and $K_2$. Define $f : K_2(X) \to \mathbb{R}$ by $f(u) := K_1(x)$ for any $x$ such that $K_2(x) = u$; this does not depend on the choice of $x$ because if $K_2(x) = K_2(y) = u$ then $x \sim y$ hence $K_1(x) = K_1(y)$. Moreover $f$ is increasing because if $u, v \in K_2(X)$ and $u < v$ then $K_2(x) = u < v = K_2(y)$ for some $x, y \in X$, hence $x \prec y$, thus $f(u) = K_1(x) < K_1(y) = f(v)$. Therefore $K_1$ and $K_2$ belong to the same equivalence class. □

This result is utility-theoretic in nature, but initial searches of the utility literature have not found a full proof (nor even a full statement), so we have provided these, along with implications for AMM completeness. The AMM literature includes treatments of some types of CFMM operations expressed in terms of keeping "utility constant" (as introduced by [10] in the context of prediction markets). We contribute to the utility-AMM interplay by connecting utility *representation* theory to the *triad* of invariance-arbitrage-completeness, in a universal framework for market systems and for arbitrage, with robust support for practical applications including fees and liquidity operations.

**Remark 5.2.** Theorem 5.1 implies that typical fee-collecting CFMMs – which are incomplete, as previously discussed – admit multiple invariants. Irrespective of completeness, the zeroth and first fundamental theorem link invariants to AMMs and arbitrage in general. Indeed, our general framework is built to handle the partial orders induced by incomplete markets that arise in practice.

# 6 Alternative notions of invariants

## 6.1 Weak invariants

*Weak* invariants have less capability than invariants to recover a DeFi system.

All invariants are weak invariants, as defined below; but not all weak invariants are invariants.

**Definition 6.1.** Let $\precsim$ be the preorder induced by a market system $M$ on $X$. If there exists $W : X \to \mathbb{R}$ satisfying

$$W(x) \leq W(y) \quad \text{for all } x, y \in X \text{ such that } x \precsim y, \tag{6.1}$$

then we say that $W$ is a *weak invariant* for $M$.

To illustrate the limitations of weak invariants compared to invariants, define an AMM $M$ by

$$X := \{(x_1, x_2) \in \mathbb{R}_+^2 : x_1 + x_2 = 1\} \text{ and } M(x) := \{y \in X : y_1 > x_1\} \text{ for } x \in X. \tag{6.2}$$

Then $M$ is complete and has invariant $K(x) = x_1$. By Theorem 5.1, the invariant is unique, and by Proposition 2.11 the invariant recovers $M$. Thus invariant $K$ uniquely provides an unambiguous description of exactly the transactions allowed by $M$. However, $M$ has multiple *weak* invariants, for instance $W(x) := 1$ for all $x \in X$. Weak invariant $W$ fails to distinguish between valid transactions such as $(0, 1) \to (1, 0)$ and invalid ones such as $(1, 0) \to (0, 1)$. A weak invariant, if strictly D-increasing, suffices to imply no-D-arbitrage, but does not suffice to recover a market system.

## 6.2 Multi-invariants

*Multi*-invariants, based on multi-utility theory [13], have greater capability to recover a DeFi system – at the cost of introducing more complexity into the invariant structure.

Let $\mathbb{R}^X$ denote the set of functions $X \to \mathbb{R}$.

For $\mathcal{K} \subseteq \mathbb{R}^X$ and $x, y \in X$, write $\mathcal{K}(x) \leq \mathcal{K}(y)$ if $K(x) \leq K(y)$ for all $K \in \mathcal{K}$.

Write $\mathcal{K}(x) < \mathcal{K}(y)$ if $\mathcal{K}(x) \leq \mathcal{K}(y)$ and $K(x) < K(y)$ for at least one $K \in \mathcal{K}$.

**Definition 6.2.** Let $\precsim$ be the preorder induced by a market system $M$ on $X$. If there exists nonempty $\mathcal{K} \subseteq \mathbb{R}^X$ such that for all $x, y \in X$ we have

$$x \precsim y \text{ if and only if } \mathcal{K}(x) \leq \mathcal{K}(y), \tag{6.3}$$

then we say that $M$ has (and admits recovery by) multi-invariant $\mathcal{K}$, and that $M$ is a *multi-CFMM*, and that $\mathcal{K}$ is a *multi-invariant* of $M$ (which recovers $M$).

The if-and-only-if in (6.3) embeds the recovery property into the definition of multi-invariant. An equivalent formulation of (6.3) is that $x \prec y$ iff $\mathcal{K}(x) < \mathcal{K}(y)$, and $x \sim y$ iff $\mathcal{K}(x) = \mathcal{K}(y)$.

**Proposition 6.3.** *Every market system $M$ has a multi-invariant (which recovers $M$).*

*Proof.* Proposition 1 of [13] adapts directly from multi-utilities to multi-invariants. □

Recall that every continuous market system is a CFMM; if the CFMM is complete, then it is recoverable by its invariant. Proposition 6.3 shows that every market system is a multi-CFMM. Irrespective of completeness, the market system is recoverable via (6.3) by its multi-invariant.

**Definition 6.4.** A multi-invariant $\mathcal{K}$ is said to be strictly D-increasing (abbreviated as "increasing") if for all $y$ D $x$ we have $\mathcal{K}(x) < \mathcal{K}(y)$. In that case we say the market system is a D-*multi-CFMM*.

The first fundamental theorem can be reformulated with an increasing multi-invariant in place of the increasing invariant, and without the continuity assumption:

**Theorem 6.5** (First fundamental theorem, multi-invariant version)**.** *A giftable market system is* D-*arbitrage-free if and only if it is a* D-*multi-CFMM.*

*Proof.* To prove "if", let $\mathcal{K}$ be an increasing multi-invariant. For any $x, y$ with $x \precsim y$, we have $\mathcal{K}(x) \leq \mathcal{K}(y)$. Therefore it cannot be the case that $x$ D $y$. Conversely, let $M$ be arbitrage-free with multi-invariant $\mathcal{K}$. For any $x, y \in X$ with $y$ D $x$, we have $x \precsim y$ by giftability. By no-arbitrage, $x \prec y$, hence $\mathcal{K}(x) < \mathcal{K}(y)$. Thus $\mathcal{K}$ is strictly D-increasing, and $M$ is a D-multi-CFMM. □

Remark 4.1 therefore extends as follows: To design an arbitrage-free DeFi system (of any type, not necessarily a DEX), the designer can restrict attention to either D-multi-CFMMs *or* D-CFMMs, because either alternative comprises precisely the arbitrage-free systems. Multi-CFMMs ensure the recovery property, but sacrifice parsimony in some cases, as a multi-invariant may have to comprise many functions (possibly infinitely many, in some cases with infinite state spaces); thus multi-CFMMs present capability/complexity tradeoffs versus (plain) CFMMs.

# 7 Conclusion

We close by extending Theorem 3.12 to include weak invariants, and by expanding Theorem 5.1 to include Proposition 2.11.

**Theorem 7.1** (First fundamental theorem)**.** *Let $M$ be a DeFi system. The following are equivalent:*

*(a) $M$ is arbitrage-free.*

*(b) $M$ has a strictly increasing invariant.*

*(c) $M$ has a strictly increasing weak invariant.*

*Proof.* We have $(a) \Rightarrow (b)$ by Theorem 3.11; and $(b) \Rightarrow (c)$ because any invariant is a weak invariant; and $(c) \Rightarrow (a)$ because weak invariance (6.1) suffices for the proof of Theorem 3.1. □

The notion of *arbitrage* has full flexibility; it may reference any strict partial order on the state space, as long as the invariant's monotonicity is understood in the same sense.

The final member of the arbitrage-invariant-completeness triad returns in the following form.

**Theorem 7.2** (Second fundamental theorem)**.** *Let $M$ be a CFMM. The following are equivalent:*

*(a) $M$ is complete.*

*(b) $M$ has an invariant that recovers $M$.*

*(c) Each invariant of $M$ recovers $M$.*

*(d) $M$ has only one invariant (modulo composition with strictly increasing mappings).*

*Proof.* Theorem 5.1 proves that $(a) \Leftrightarrow (d)$. Proposition 2.11 proves that $(b) \Rightarrow (a) \Rightarrow (c)$. As $M$ has an invariant, $(c) \Rightarrow (b)$, completing the chain. □

The second fundamental theorem applies to all continuous market systems, including all AMMs, because all of them are CFMMs, according to the zeroth fundamental theorem.